\documentclass[12pt]{article}

\usepackage{authblk}

\title{
Supporting 64-bit global indices in Epetra and other Trilinos packages --
Techniques used and lessons learned
}
\author[1]{Chetan~Jhurani\thanks{chetan.jhurani@gmail.com, corresponding author}}
\author[1]{Travis~M.~Austin\thanks{austint73@gmail.com, now at MSC Software}}
\author[2]{Michael~A.~Heroux\thanks{maherou@sandia.gov}}
\author[2]{James~M.~Willenbring\thanks{jmwille@sandia.gov}}
\affil[1]{Tech-X Corporation, 5621 Arapahoe Ave, Boulder, Colorado 80303, U.S.A.}
\affil[2]{Sandia National Laboratories, Albuquerque, New Mexico 87185, U.S.A.}

\newcommand{\Sec}[1] {Section~\ref{#1}}

\begin{document}

\maketitle

\begin{abstract}

The Trilinos Project is an effort to facilitate the design,
development, integration and ongoing support of mathematical software
libraries within an object-oriented framework.  It is intended for
large-scale, complex multiphysics engineering and scientific
applications~\cite{Trilinos-Overview,Trilinos-Users-Guide,trilinos_toms}.
Epetra is one of its basic packages. It provides serial and parallel
linear algebra capabilities.  Before Trilinos version 11.0, released
in 2012, Epetra used the C++ int data-type for storing global and
local indices for degrees of freedom (DOFs).  Since int is
typically 32-bit, this limited the largest problem size to be smaller
than approximately two billion DOFs.  This was true even if a
distributed memory machine could handle larger problems.  We have
added optional support for C++ long long data-type, which is at least
64-bit wide, for global indices.  To save memory, maintain the speed
of memory-bound operations, and reduce further changes to the code,
the local indices are still 32-bit.
We document the changes required to achieve this feature and how
the new functionality can be used.  We also
report on the lessons learned in modifying a mature and popular
package from various perspectives -- design goals, backward
compatibility, engineering decisions, C++ language features, effects
on existing users and other packages, and build integration.

\end{abstract}




\section{Introduction}

Large scientific and engineering simulations are typically run in
parallel on machines with distributed memory.  Recent progress in
distributed hardware has made it possible to run simulations with
number of degrees of freedom (DOFs) that cannot fit the C++ int
data-type.  Thus, it becomes necessary to modify existing codes or
create new ones so that hardware can be utilized fully for more
accurate and reliable simulations.

Trilinos~\cite{Trilinos-Overview,Trilinos-Users-Guide,trilinos_toms}
is a popular and mature software library that serves as a foundation of
various applications.  It provides linear algebra data structures like
vectors, matrices, maps, and also algorithms that act on them.  Till
recently, the only way of using 64-bit global indices in Trilinos was
to use its newer package Tpetra, which is C++ template based.  This
allowed applications to break the barrier of 2 billion DOFs (or
$2^{31} - 1$ DOFs to be precise).  However, many Trilinos packages and
end-user codes depend on the older Epetra package that till recently
did not have 64-bit support.

We have added support for C++ long long data-type which is at least
64-bit wide (and typically is exactly 64-bit wide).  It can be used for
storing global indices in Epetra.  To save memory, maintain the speed
of memory-bound operations, and reduce further changes to the code,
the local indices are still int-based, typically 32-bit.  Before our modifications, Epetra used
int data-type for both global and local indices, without any typedefs,
and this design assumption was deeply ingrained.  The type int was also used for
data that is not a global or local index, and thus it is not possible
to easily search and replace with a typedef.  Even if this was
possible within Epetra, such a change would imply that a lot of user
code would also have to use a typedef for compatibility.  Instead of
typedef or template mechanisms, which are completely compile-time
features, we have used the following combination of C++
language~\cite{Stroustrup2000} features and preprocessor and build
tool features to retrofit Epetra.
\begin{itemize}
\item C++ function-overloading
\item type-conversion for function return values so as to implicitly convert
a long long to int where it can fit without loss of data.
\item C++ exceptions when a ``wrong'' function is called for a
data-type
\item Preprocessor and CMake~\cite{Martin2003} flags to enforce correctness
at compile-time
\item A judicious and minimal use of C++ templates internally to avoid
code duplication
\end{itemize}

There are significant differences between how we added this 64-bit support
to Epetra (and some other dependent Trilinos packages), and how
such a feature is made available in Tpetra and PETSc~\cite{Balay2013}.
Although much of this article is on Epetra and dependent packages, we
briefly mention how this support is present in other packages.  Tpetra
relies on C++ templates and compile-time specification of both local
and global index types.  A user can choose the desired index type when
constructing and using Tpetra objects.  Different objects in the same
application can have different index types at the same time.  PETSc uses the typedef
feature to handle types of indices.  When building PETSc, one specifies whether
a PetscInt typedef should be 64-bit using the option
{\tt --with-64-bit-indices}.  By default it is 32-bit.  Since PETSc
consistently uses PetscInt for indices, and expects users to do the same,
changing the typedef meaning
in just one location leads to a different type everywhere.  A drawback
is that an application can only use a single index type unless
both PETSc and the application are recompiled
for a different type.  Both these techniques are common in other softwares
and have advantages and disadvantages.  However, our goals, listed ahead, are
different and both these techniques are not suitable for modifying Epetra.

We document the changes required to achieve 64-bit compatible Epetra.
Rather than just produce only a list of what changed and how to use
the new functionality, we also present lessons learned in modifying a
mature and popular package from various perspectives -- design goals,
backward compatibility, engineering decisions, compile-time and
run-time issues, C++ language features, effects on existing users and
other packages, and build integration.

Users who wish to convert their existing Epetra-based source
to take advantage of 64-bit functionality should go through
Sections~\ref{sec:using}, \ref{sec:modes}, and \ref{sec:tests}
at least.

Here is an outline of the sections ahead.  We briefly mention the
basic Epetra data structures for new users in~\Sec{sec:basicds}.
\Sec{sec:goals} lists our design goal for this task.
\Sec{sec:status} reports the high-level progress and future tasks.
In~\Sec{sec:using}, we describe the changes in detail and this will
be useful for users who wish to use the 64-bit functionality.
\Sec{sec:dev} contains information for developers.
\Sec{sec:modes} shows how to switch off selected functionality (32-bit or
64-bit) at compile-time.  This is useful for debugging and initial
porting.  Existing users should pay attention to
backward compatibility issues we describe in~\Sec{sec:backward}.
In~\Sec{sec:dependentpack}, we describe the changes made to two
other Trilinos packages -- AztecOO and TriUtils.
Section~\ref{sec:tests} shows how we test the new functionality.
Finally, in~\Sec{sec:lessons} we list a few facts that might help
people who may want to retrofit a software package for different data types.

\section{Basic Epetra data structures}
\label{sec:basicds}

Many linear algebra Epetra classes, serial and distributed, use one or
more instances of the Epetra\_BlockMap class directly or indirectly.  This class manages the
global sizes and layouts of the objects with global indices.  Since
this class is limited by the size of C++ int, other Epetra
classes, for example for graphs, matrices, vectors, and Epetra-dependent
Trilinos packages also have this limitation.  The goal is to modify
basic Epetra classes like Epetra\_BlockMap, Epetra\_CrsGraph,
Epetra\_CrsMatrix, Epetra\_Vector, and many other classes so they
contain data structures and member functions that provide them with
64-bit capability.

\section{Design goals}
\label{sec:goals}

Here are our design goals in removing the 32-bit global index
limitation.  This was not our full list of goals when we started the
project in late 2011.  But as we experimented and learned more over
time, this detailed list came into existence.
\begin{enumerate}
\item Remove the size limitation by using long long for global sizes in
appropriate locations.
\item Preserve int for local sizes and local indices so that any
increase in the storage and run-time for memory-bound routines is
minimal.
\item Trilinos packages that depend on Epetra should not be heavily
affected.
\item Ideally, the code of existing users should compile without making any
changes.  A few harmless warnings are all right.
\item There should be little to no effect on run-time performance for
existing (32-bit) users.
\item Users who wish to use the 64-bit functionality would have to make very
small changes to their existing int based code.  Ideally it should
be possible to find a large fraction of changes at compile-time.
\item Very little code should be duplicated for the two distinct data
types, using C++ templates locally for example.
\item C++ templates are not to be exposed in any publicly visible
function and only be used internally.
\item New testing code should be created to test the long long
functionality.
\item It should be possible to ``switch off'' 64-bit functionality so
that complete backward compatibility is maintained
\item It should be possible to ``switch off'' 32-bit functionality so
that new users can find locations where their code is not 64-bit
compatible.
\end{enumerate}
Achieving the last two goals would allow users to check at
compile-time whether their code is incompatible with 32-bit Epetra or
64-bit Epetra or both.  We will show ahead how all these goals are
achieved.

\section{Status of the transition}
\label{sec:status}

The 64-bit addition to Epetra and dependent packages is an ongoing
effort.  We have modified Epetra nearly completely except for a few
classes that are not used very often.  We mention their names for the
sake of completeness.  They are
CrsSingletonFilter,
FEVbrMatrix,
LinearProblemRedistor,
MapColoring,
RowMatrixTransposer, and
VbrMatrix
each with the prefix Epetra\_.

Many other packages in Trilinos depend on Epetra.  We have currently
modified TriUtils (utilities for other Trilinos packages) and AztecOO
(preconditioned Krylov methods) so they are 64-bit compatible.  We are
working on Ifpack (algebraic preconditioners) and EpetraExt
(construction and service functions for linear algebra).

\section{Using the 64-bit functionality}
\label{sec:using}

We now describe how to use the 64-bit functionality starting from map
construction and then moving on to member functions of other important
classes for vectors, graphs, and matrices.  The starting point in
using the 64-bit support is that the user should construct 64-bit maps
using the long long based constructors.  Such maps in turn lead vectors,
graphs, and matrices that are 64-bit enabled. The user can then call
new functions in these data structure classes that support 64-bit
input and output.  We will list the specific functions.

\subsection{Map construction}

The code below shows a common way of creating a 32-bit map with a
given number of global indices.  The constructor then divides the
elements evenly across all processors in the communicator.
\begin{verbatim}
  Epetra_MpiComm Comm(MPI_COMM_WORLD);
  int NumGlobalElements = 10;
  int IndexBase = 0;
  Epetra_Map Map(NumGlobalElements, IndexBase, Comm);
\end{verbatim}
Another way is to create a map that puts NumMyElements global indices
on the calling processor. The array MyGlobalElements, which passes the
global indices, is also provided.  If NumGlobalElements is -1, the
number of global elements are computed by the constructor.  Here is
its signature.
\begin{verbatim}
  Epetra_Map(int NumGlobalElements, int NumMyElements,
             const int *MyGlobalElements,
             int IndexBase, const Epetra_Comm& Comm);
\end{verbatim}

To use 64-bit functionality, the caller must use long long data-type for
certain arguments.  We provide extra constructors for that.  For
example, the first constructor call becomes
\begin{verbatim}
  long long NumGlobalElements = 10;
  int IndexBase = 0; // IndexBase can be int or long long
  Epetra_Map Map(NumGlobalElements, IndexBase, Comm);
\end{verbatim}
Similarly, the long long version of the second constructor is
\begin{verbatim}
  Epetra_Map(long long NumGlobalElements, int NumMyElements,
             const long long *MyGlobalElements,
             int IndexBase, const Epetra_Comm& Comm);
\end{verbatim}
As before, IndexBase can be long long or int for
this constructor.

Note that NumMyElements is still an int,
which means the number of local indices (indices on a single
processor) is still limited by 32-bit integer type.  This is so that
one does not increase local index storage and reduce performance of
memory-bound operations just because global indices are 64-bit.  This
is a design assumption keeping in mind the typical use-case and
typical local memory sizes.  Even if one were to change local
index type to long long, the number of changes would be much higher.

In summary, every map constructor has overloads for two different
integral types.  Similar changes are present in the low-level class
Epetra\_BlockMap.  A map then has the knowledge of which kind of
map it is -- 32-bit or 64-bit.  This construction process also allows
user to choose the data-type at run-time.  No change is required to
function signatures that may have (previously constructed) maps as
arguments.

When passing a literal -1 for the first argument (NumGlobalElements),
one should do it like (long long)-1 or -1LL or explicitly use a named
long long variable.  This holds for other function calls also where a
literal is passed without a named variable and remaining arguments are
not sufficient to deduce that a long long based function should be
called.  Otherwise, the int overload will be called.

\subsection{32-bit and 64-bit state of maps}

Maps can be in either of two states -- 32-bit or 64-bit -- or none of
them in case they were default constructed.  Thus, it is important to
query the state.  The following new functions are added to
Epetra\_BlockMap for this requirement.

\begin{verbatim}
  // Returns true if map created with int
  bool GlobalIndicesInt() const;
  
  // Returns true if map create with long long
  bool GlobalIndicesLongLong() const;
  
  // Returns true if map index type is of type int_type
  // Useful to get type at compile-type via templates
  template<typename int_type>
  bool GlobalIndicesIsType() const;
  
  // Returns true if the global index type is valid
  bool GlobalIndicesTypeValid() const;
  
  // Returns true if the global index types match
  bool GlobalIndicesTypeMatch(
    const Epetra_BlockMap& other) const;
\end{verbatim}

This is a good place to mention another important design assumption.
We expect that objects with identical global index type will interact
with each other.  For example, we do not allow a matrix with int based
row map and long long based column map.  Such conditions can be enforced
at run-time only and we throw an exception whenever we encounter
mixing of int and long long based objects.  We use GlobalIndicesTypeMatch
function shown above for checking this.

\subsection{GID related map member functions}
\label{sec:suffix}

GID (global index) related functions in map classes typically have an
int return type.  Since C++ does not have overloading by return
type, we cannot create functions of the same name and just change the
return type.  To maintain backward compatibility, new functions have
been added with a suffix ``64''.  The functions that could be
differentiated based on argument data type were not given a suffix.
For example, we have
\begin{verbatim}
  // Return GID for int based maps
  // Exception for long long based maps
  int GID(int LID) const;
  
  // Return GID for int or long long based maps
  long long GID64(int LID) const;
\end{verbatim}
The idea is that any old user code that used GID(...) will continue to
work as usual but a new code that is forward-looking will use
GID64(...) since it works for both kinds of maps.  Since GID(...)
might lead to overflow in case of long long map we are extra cautious
and throw an exception in all cases, even if the return value might
fit an int.

Similar changes are made to other int returning functions.
Here is the list of the new suffixed functions.
\begin{verbatim}
  long long MinAllGID64() const;
  long long MaxAllGID64() const;
  long long MinMyGID64() const;
  long long MaxMyGID64() const;
  long long NumGlobalElements64() const;
  long long IndexBase64() const;
  long long NumGlobalPoints64() const;
\end{verbatim}

There are other functions that do not return values but return
pointers instead.  In these cases, the long long one cannot work
correctly for int based maps and is a mistake.  Thus the right way
is to throw an exception.
\begin{verbatim}
  // Old function.  Exception for long long based maps.
  int* MyGlobalElements() const;
  
  // Additional function.  Exception for int based maps.
  long long* MyGlobalElements64() const;
\end{verbatim}

There also exists code that did not receive global indices by a
returned pointer but it copied the indices to a user provided pointer.
This is easily handled by overloading and there is no need for a
suffix.
\begin{verbatim}
  int MyGlobalElements(int * MyGlobalElementList) const;
  int MyGlobalElements(long long * MyGlobalElementList) const;
\end{verbatim}

Sometimes, however, it is easier to not worry about separate calls
(MyGlobalElements and MyGlobalElements64) when returning pointers.  It
is easier to get two pointers and pick the one that is non-zero.  This
also simplifies switching-off 32-bit or 64-bit functions as will show
in~\Sec{sec:cmake}.  The following two functions are added for this
purpose.  One of the pointers in the arguments will be zero after
calling the function.
\begin{verbatim}
  void MyGlobalElements(const int*& IntGIDs,
                        const long long*& LLGIDs) const;
                        
  void MyGlobalElements(int*& IntGIDs,
                        long long*& LLGIDs);
\end{verbatim}

Computing an LID (local index) for a given GID is done by the LID
function.  We overload it for long long.
\begin{verbatim}
  // Exception for long long based map
  int LID(int GID_in) const;
  
  // Works for int and long long based maps.
  int LID(long long GID_in) const;
\end{verbatim}
Similarly, checking if a given GID is local is done by the MyGID
function.  We overload it for long long.
\begin{verbatim}
  // Exception for long long based map
  bool MyGID(int GID_in) const;
  
  // Works for int and long long based maps.
  bool MyGID(long long GID_in) const;
\end{verbatim}

\subsection{Additions to the row matrix interface}
\label{sec:rowmatrix}

The Epetra\_RowMatrix is an abstract class with all pure virtual
functions.  The interface is intended for real-valued double-precision
row-oriented sparse matrices.  It is implemented by Epetra\_CrsMatrix
and many classes in other packages such as Ifpack, ML, EpetraExt, and
AztecOO.  Since it specifies some functions that return int for
GIDs, the return type should be long long for long long based matrices.  As
before we add new functions with suffix 64 to the interface.  Since
they are pure virtual functions, they have to be implemented by all
deriving concrete classes.  We have added such implementations for
classes in Trilinos.

Here are the original four functions.
\begin{verbatim}
    virtual int NumGlobalNonzeros() const = 0;
    virtual int NumGlobalRows() const = 0;
    virtual int NumGlobalCols() const = 0;
    virtual int NumGlobalDiagonals() const = 0;
\end{verbatim}
The added functions are
\begin{verbatim}
    virtual long long NumGlobalNonzeros64() const = 0;
    virtual long long NumGlobalRows64() const = 0;
    virtual long long NumGlobalCols64() const = 0;
    virtual long long NumGlobalDiagonals64() const = 0;
\end{verbatim}
The logic remains the same.  Any old user code that used the
non-suffixed version will continue to work as usual but a new code
that is forward-looking will use the suffixed version.  A user
implementing this abstract class should implement these functions so
that the new functions work for both kinds of maps and the
non-suffixed version should throw an exception if called for long long
based matrix.

\subsection{New classes for long long global indices}

For brevity, we ignore the Epetra\_ prefix in class names here.
Epetra contains these three classes that store int data --
IntVector, IntSerialDenseVector, and
IntSerialDenseMatrix.  They are also used for storing global
indices.  Introduction of long long global indices required objects that
can store long long data.  This facility is provided by the following
three new classes -- LongLongVector,
LongLongSerialDenseVector, and
LongLongSerialDenseMatrix.  The interface is identical up to
the change that in certain locations the int keyword has been
replaced with long long.

IntSerialDenseVector is used as function argument in
FECrsMatrix class.  The corresponding overloaded functions for
long long use LongLongSerialDenseVector.  This pattern can be
used in other such situations.

Additionally, to allow compile-time
polymorphism using templates, two new classes have been added --
GIDTypeVector and GIDTypeSerialDenseVector.
GIDTypeVector\textless int\textgreater :: impl is a typedef for
IntVector and GIDTypeVector\textless long
long\textgreater :: impl is a typedef for LongLongVector.
Hence, inside template functions, one can write code that works for
both int and long long.

\subsection{Vector, Graph, and Matrix construction}

The classes Epetra\_MultiVector, Epetra\_Vector, Epetra\_CrsGraph, and
Epetra\_CrsMatrix are a few of the basic and most important in Epetra.
All of them take an Epetra\_BlockMap or
Epetra\_Map argument in their constructors.  This keeps the 32-bit or
64-bit issue hidden and thus their construction process does not have
to change.  The additions are only for some of the member functions.
They are those that have global indices as arguments or return types.
We describe this ahead.

Just like for maps, a suffixed function returns a long long and thus it
can meaningfully return an int value.  It will work for both
32-bit and 64-bit objects.  The corresponding non-suffixed function
will throw and exception if called for 32-bit objects.

\subsection{MultiVector and Vector member functions}

We show the new functions added to the class Epetra\_MultiVector and
Epetra\_Vector, its derived class.  Similar changes have been made to
Epetra\_FEVector, the derived class for Finite Elements, but we do not
show them here to reduce duplication.
In Epetra\_MultiVector, the functions ReplaceGlobalValue and
SumIntoGlobalValue are overloaded so that the new functions take a
long long global row index.  To return the global length we need a new
function with the suffix 64, as explained in~\Sec{sec:suffix}.  Its
signature follows.
\begin{verbatim}
  long long GlobalLength64() const;
\end{verbatim}

Epetra\_Vector has two new overloaded member functions for working
with 64-bit indices -- ReplaceGlobalValues and ChangeValues.  Both the
new functions take pointers to long long instead of pointers to int.

\subsection{Graph and Matrix member functions}

There are many additional member functions in Epetra\_CrsGraph and
Epetra\_CrsMatrix to support long long functionality.  Rather than show
their new signatures, we just list the names of overloaded or suffixed
functions.  As before, the functions with suffix 64 are new functions
that return a long long and those without the suffix are overloads with
long long arguments instead of  int.

Both classes implement the four functions of Epetra\_RowMatrix
mentioned earlier.  These are NumGlobalNonzeros64, NumGlobalRows64,
NumGlobalCols64, and NumGlobalDiagonals64. See~\Sec{sec:rowmatrix}
for further details.  They also implement IndexBase64, GRID64, GCID64,
MyGRID, MyGCID, LRID, LCID, and MyGlobalRow.

In addition to these, the following new functions in Epetra\_CrsGraph
return long long and thus need the 64 suffix.  They are
NumGlobalBlockRows64, NumGlobalBlockCols64, NumGlobalBlockDiagonals64,
and NumGlobalEntries64.

We have also added the overloaded counterparts for
InsertGlobalIndices, RemoveGlobalIndices, ExtractGlobalRowCopy,
ExtractGlobalRowView, RemoveGlobalIndices, NumGlobalIndices,
NumAllocatedGlobalIndices in Epetra\_CrsGraph.  Their global index
related arguments are long long instead of int.

Epetra\_CrsMatrix class overloads these functions for long long instead
of int -- InsertGlobalValues, InsertGlobalValues,
ReplaceGlobalValues, SumIntoGlobalValues, ExtractGlobalRowCopy,
ExtractGlobalRowView, and NumGlobalEntries.

Similar additions have been made to Epetra\_FECrsGraph and
Epetra\_FECrsMatrix, the derived classes for Finite Elements, but we
do not show them here to reduce duplication.

\section{Information for developers}
\label{sec:dev}

This section briefly describes internal implementation details.
Trilinos users can skip this part.

\subsection{Changes to CrsGraphData}

We have changed the struct EntriesInOneRow nested in
the class Epetra\_CrsGraphData to be dependent on a template argument int\_type.
The class is instantiated with an int and a long long.  This is done
because a user who fills in the CrsGraph objects can do it using
global indices (int or long long) or local indices (always int).
Since this is a run-time decision by caller and not a compile time
decision, Epetra\_CrsGraphData uses instantiation with both types but
populates the one that is necessary.

We have also defined a new nested struct IndexData in
Epetra\_CrsGraphData.  It is dependent on a template parameter
int\_type and contains all the index related data.  One can get the
reference to the appropriate data with the functions
\begin{verbatim}
  Epetra_CrsGraphData::IndexData<long long>&
    Epetra_CrsGraphData::Data<long long>();
\end{verbatim}
and
\begin{verbatim}
  Epetra_CrsGraphData::IndexData<int>&
    Epetra_CrsGraphData::Data<int>();
\end{verbatim}
The first function will work only if the map's global indices are
long long based and indices have not been made local (by a call to
MakeIndicesLocal, for example).  The second function will work if the
map's global indices are int based or they are long long based but
have been made local.  If the conditions are not satisfied, we throw
an exception.

\subsection{Wrappers over MPI collectives}

We have added the long long versions of the following MPI collective
functions $-$ GatherAll, SumAll, MaxAll, MinAll, ScanSum.  They have
been added to the classes Epetra\_Comm, Epetra\_SerialComm, and
Epetra\_MpiComm.

\subsection{Companion arrays in sorting}

The function Epetra\_Util::Sort is overloaded for long long in two ways.
One addition takes long long companion arrays for simultaneous sorting.
Another addition allows for long long keys.  We show both of them.
Just like for other companion arrays, if there is no long long companion,
one can pass zero for number of companions and a null pointer.
\begin{verbatim}
void Sort(
  bool SortAscending, int NumKeys, int* Keys, 
  int NumDoubleCompanions, double ** DoubleCompanions, 
  int NumIntCompanions, int ** IntCompanions,
  int NumLongLongCompanions, long long ** LongLongCompanions);

void Sort(
  bool SortAscending, int NumKeys, long long* Keys, 
  int NumDoubleCompanions, double ** DoubleCompanions, 
  int NumIntCompanions, int ** IntCompanions,
  int NumLongLongCompanions, long long ** LongLongCompanions);
\end{verbatim}

\subsection{Additions to Distributor and Directory classes}

The Epetra\_Distributor class is an interface that encapsulates gather
and scatter operations on parallel hardware.  It has two
implementations --  Epetra\_SerialDistributor and
Epetra\_MpiDistributor.  All these now have an additional member
function overloaded for long long data.
\begin{verbatim}
int CreateFromRecvs(
  const int & NumRemoteIDs, const long long * RemoteGIDs,
  const int * RemotePIDs, bool Deterministic,
  int & NumExportIDs, long long *& ExportGIDs,
  int *& ExportPIDs);
\end{verbatim}

The Epetra\_Directory class is an abstract base class to reference
non-local elements.  Epetra\_BasicDirectory is its implementation.
Both classes now have the following new overloaded function.

\begin{verbatim}
int GetDirectoryEntries(
  const Epetra_BlockMap& Map, const int NumEntries,
  const long long * GlobalEntries, int * Procs,
  int * LocalEntries, int * EntrySizes,
  bool high_rank_sharing_procs=false) const;
\end{verbatim}

\section{Multiple compilation modes}
\label{sec:modes}

Trilinos now has three distinct modes of compilation for global index
types that can be chosen using compile time flags.
See~\Sec{sec:cmake} for how these can be specified in CMake.  The
three modes are as follows.
\begin{enumerate}
\item 32-bit and 64-bit. This is the default mode. In this mode both 32-bit and
64-bit global index code coexist so the user can choose the best data
size for global indices at run-time.
\item 32-bit. This mode hides all the new 64-bit interfaces and
implementation.  The objective is to give users the old interface for
full compatibility.
\item 64-bit. This mode hides the old 32-bit interfaces and
implementations that are counterparts to the new 64-bit interfaces.
The objective is to ensure that new 64-bit code is not relying on any
32-bit code internally.  This helps in debugging and porting code that
desires new functionality.
\end{enumerate}

The dual mode is backward compatible up to the extent that user code
does not rely on automatic type conversion (see~\Sec{sec:conversion})
and user had not implemented Epetra\_RowMatrix interface
(see~\Sec{sec:rowmatrix}).  If the user had implemented it, four
additional functions must be implemented so that the class is not
abstract.

Here is an example from Epetra\_BlockMap header of the two kinds of
functions and how they exist in different modes.

\begin{verbatim}
#ifndef EPETRA_NO_32BIT_GLOBAL_INDICES
  int LID(int GID) const;
#endif
#ifndef EPETRA_NO_64BIT_GLOBAL_INDICES
  int LID(long long GID) const;
#endif
\end{verbatim}
\begin{verbatim}
#ifndef EPETRA_NO_32BIT_GLOBAL_INDICES
  int GID(int LID) const; 
#endif
  long long GID64(int LID) const; 
\end{verbatim}
Note that GID64 is present in all modes and its return value does not
depend on the run-time chosen object mode.  This makes it simple to
use in downstream packages irrespective of the compile-time mode.
See~\Sec{sec:suffix} for further details.

The preprocessor macros don't have to be defined by the user but are
supplied (if at all) by the CMake build system.  In the dual mode both
\begin{itemize}
	\item EPETRA\_NO\_32BIT\_GLOBAL\_INDICES and
	\item EPETRA\_NO\_64BIT\_GLOBAL\_INDICES
\end{itemize}
are left undefined.  The first one is
defined in pure 64-bit mode and the second one is defined in pure
32-bit mode.  A code that wishes to use these macros must include
the Epetra\_ConfigDefs.h header file.

Any user code for the new 64-bit functionality should ideally compile
for all three modes preferably with few or no conversion warnings.
This ensures that a large fraction of porting effort is complete.

\subsection{Integration with CMake}
\label{sec:cmake}
The preprocessor macros discussed earlier can be controlled via options given
to CMake at configuration time. The corresponding CMake macros are
\begin{itemize}
	\item Trilinos\_NO\_32BIT\_GLOBAL\_INDICES and
	\item Trilinos\_NO\_64BIT\_GLOBAL\_INDICES
\end{itemize}

\subsection{Multiple mode design pattern}

We describe a design pattern so that one can write code appropriate
for multiple modes and one that has minimal run-time overhead.
Consider a function that wishes to use an Epetra\_Map, which may
be 32-bit or 64-bit.  It can be written like this and it works
for all three compilation modes.  Typically such a code is
needed only in a few critical locations.
\begin{verbatim}
void func(const Epetra_Map& Map)
{
#ifndef EPETRA_NO_32BIT_GLOBAL_INDICES
  if(Map.GlobalIndicesInt()) {
    const int* gids = Map.MyGlobalElements();
    // Use gids now
  }
  else
#endif
#ifndef EPETRA_NO_64BIT_GLOBAL_INDICES
  if(Map.GlobalIndicesLongLong()) {
    const long long* gids = Map.MyGlobalElements64();
    // Use gids now
  }
  else
#endif
    throw "func: GlobalIndices type unknown";
}
\end{verbatim}
Note that instead of duplicating code manually after ``Use gids now'',
one may also call a C++ template function with template
arguments long long or int.

Since these specific functions (MyGlobalElements and its 64-bit version)
are very common, we have provided a new function specifically
for them so that such verbosity and use of macros can be avoided.  See~\Sec{sec:suffix}
for details.  However, in other cases, this pattern is very useful.

\section{Backward compatibility}
\label{sec:backward}

We mention two cases where existing user code (written for older 32-bit Epetra)
can fail to compile when 64-bit additions are also present.

\subsection{Ambiguity in type conversion}
\label{sec:conversion}

The addition of new constructors and other functions uses the C++
function overloading mechanism.  This is backward compatible to the
extent that automatic type conversion are not expected.  For example,
before we added the new constructors, any user code that relied on
conversion from types other than int, say unsigned int or short or
size\_t relied on type conversion in passing by value.  Such code will
not compile when both int and long long versions are present because
of the resulting ambiguity.  The best way is to not rely on type
conversion.  However, in the short term, one may switch-off the 64-bit
functionality using CMake flags discussed in~\Sec{sec:cmake}.
This will remove the ambiguity and the errors due to it.  

\subsection{New pure virtual functions}

We have added new pure virtual functions to Epetra\_RowMatrix
interface for 64-bit functionality.  The interface is intended for
real-valued double-precision row-oriented sparse matrices.  Any
user code that implemented this
interface will now require four simple additional functions.
See~\Sec{sec:rowmatrix} for details.  Of course, if the user never
intends to use the interface for 64-bit functionality, a simple
implementation that throws an exception unconditionally will suffice.

\section{Changes to other Trilinos packages}
\label{sec:dependentpack}

We have modified the packages TriUtils (utilities for other Trilinos packages) and
AztecOO (preconditioned Krylov methods) so they are 64-bit compatible.
AztecOO does not have any publicly visible changes for 64-bit
compatibility.

Triutils now has the following new 64-bit global functions.  The ones
with suffix 64 are those that cannot be distinguished by
C++ overloading since they construct maps rather than use any
pre-existing input maps.
\begin{itemize}
\item Trilinos\_Util\_CountTriples
\item Trilinos\_Util\_CountMatrixMarket
\item Trilinos\_Util\_ReadMatrixMarket2Epetra64
\item Trilinos\_Util\_ReadHb2Epetra64
\item Trilinos\_Util\_ReadHpc2Epetra64
\item Trilinos\_Util\_GenerateCrsProblem64
\item Trilinos\_Util\_ReadTriples2Epetra64
\end{itemize}
We don't expect that users
will read really large matrices (that require 64-bit GIDs) from a file.  Still, it is useful for
testing to read a small matrix as if it required 64-bit.

Additionally, the class CrsMatrixGallery in namespace Trilinos\_Util
has a new argument ``bool UseLongLong''.  A true value can be supplied
to create a 64-bit compatible matrix.

\section{Test suite}
\label{sec:tests}

We have added around 20 new tests to Epetra by modifying the existing test suite so that the
long long functionality can be tested.  They are present in ``epetra/test''
directory and are suffixed with ``\_LL''. Most of the changes are local
one-line changes.  It is instructive to compare the new tests
with the original code to see how to convert an application from
int to long long.  Similar new tests have been added to AztecOO and
TriUtils.

These tests run regularly and help in catching regressions.
The older tests use 32-bit maps and work whether the CMake flag
\begin{quote}
Trilinos\_NO\_64BIT\_GLOBAL\_INDICES
\end{quote}
is enabled or not.  Similarly, the new \_LL tests work whether CMake flag
\begin{quote}
Trilinos\_NO\_32BIT\_GLOBAL\_INDICES
\end{quote}
is enabled or not and use 64-bit maps.  See~\Sec{sec:modes} for a discussion
of these flags.

\section{Lessons learned and guidelines}
\label{sec:lessons}

We list some lessons learned in carrying out the task of retrofitting
Epetra with 64-bit support.  The purpose of listing the lessons is two-fold.
Firstly, it might be useful for someone carrying out such a task to
other old or popular software packages with minimal disturbance to
backward compatibility.  Presumably, there would be quite a few
older packages written in days when C or C++ int was of sufficient
size and retrofitting them for a different size might be attractive.  It
could also be that one may want to go from double precision to
single precision floating point data.  Secondly, it might help someone
writing new software in understanding the effort and tools required to retrofit.
This way, perhaps, the decision of future-proofing the data types or providing tools
to choose them can be made with better information.

\subsection{Effort estimate}

Although it is hard
to predict or even precisely measure how long a task like this takes.
This is true partly because it is not done in isolation.
Still, it is useful to know a rough estimate.
We estimate that it took approximately 7 person-months spread over 15
calendar months and distributed amongst a team of 4 people (the authors) and
a few other team members who helped in testing and discussion of issues.
Specifically for Epetra, it required changing its library source in
around 800 different locations in around 70 files.  Changes to
tests, examples, and other packages are not mentioned.

\subsection{Language and compiler facilities}

As mentioned previously, we have heavily made of use function overloading,
C++ templates for internal implementation, implicit type conversion when
it does not truncate, and exceptions to prevent
execution of functions (when the wrong type function is called).
These facilities made it possible to achieve usable 64-bit functionality.
However, the two features with greatest impact on reliability and speed
of retrofitting were creating compilation modes using the preprocessor and
conversion warnings.

The first task, where low-level Epetra functions
were enclosed by {\tt \#ifdef} macros (see~\Sec{sec:modes}), was
time-consuming and repetitive but it paid off.  It helped in pointing
out, via compilation errors, higher-level code that had to be changed.
Once such locations were found, it became a matter of intelligently
choosing which language feature was appropriate to add 64-bit
functionality.  Once any new functionality was added, the older
and new code was again enclosed in {\tt \#ifdef} macros.  This
process was repeated with higher-level code till the end.
Compilation modes helped initially in retrofitting but later helped
in finding out bugs that can be found only by running a program.

Simultaneously, it was important to watch for type conversion warnings
within Trilinos.
The {\tt -Wall} option in GCC (GNU Compiler Collection) does not
include the {\tt -Wconversion} warning.  However, it is crucial
for our task and had to be enabled.  See~\cite{wconversion}
for a rationale of why it is not included in {\tt -Wall} and examples.

An issue which we had not anticipated was ambiguity due to multiple
types (see~\Sec{sec:backward}).  This would not be an issue if users passed
the exact type (int in our case).  However, many users rely on
implicit conversion usually from unsigned int. This issue
is preventable if users have removed sign conversion warnings.

\subsection{Return type}

Returning a computed value in a function using the return mechanism
is quite natural.  However, C++ does not have overloading by return type~\cite{Stroustrup2000}.
This led to addition of extra functions with a different suffix.
This made the code work, but it was not a very clean method.

\vspace{1cm}

\noindent \textbf{ \large{Acknowledgements}}

We thank Karen Devine and other first users who provided helpful bug
reports, Roscoe A. Bartlett for discussion of backward compatibility
issues, Brent M. Perschbacher for helping in source control
management and continuous integration, and Bill Spotz for working
on PyTrilinos and 64-bit related issues.

\bibliographystyle{plain}
\bibliography{epetra64}

\end{document}